\documentclass[11pt,journal]{IEEEtran}
\usepackage{setspace}
\usepackage{cite}
\usepackage[T1]{fontenc}

\usepackage[pdftex]{graphicx}
\graphicspath{/figs/}
\DeclareGraphicsExtensions{.pdf,.jpeg,.png, eps}
\usepackage{amsmath, amssymb}

\usepackage[table,xcdraw]{xcolor}
\usepackage{bm}

\usepackage{float}
\usepackage[caption = false]{subfig}
\usepackage{algorithm}
\usepackage{algorithmicx}
\usepackage{algpseudocode}
\usepackage{pifont}
\usepackage{xcolor}

\usepackage{array}
\usepackage{url}
\usepackage{parskip}
\let\OLDthebibliography\thebibliography
\renewcommand\thebibliography[1]{
	\OLDthebibliography{#1}
	\setlength{\parskip}{0pt}
	\setlength{\itemsep}{2pt plus 0.1ex}
}
\renewenvironment{IEEEbiography}[1]
{\IEEEbiographynophoto{#1}}
{\endIEEEbiographynophoto}

\usepackage{mathtools}

\algdef{SE}[DOWHILE]{Do}{doWhile}{\algorithmicdo}[1]{\algorithmicwhile\ #1}%

\newcommand{\Lcsi}{L_{\text{CSI}}}

\begin{document}
	
	\title{Massive MIMO in 5G: How Beamforming, Codebooks, and Feedback Enable Larger Arrays}
	\author{\IEEEauthorblockN{Ryan M. Dreifuerst,~\IEEEmembership{Student Member,~IEEE, }}%
		\and
		\IEEEauthorblockN{Robert W. Heath Jr.~\IEEEmembership{Fellow,~IEEE, }}%
		\thanks{Ryan M. Dreifuerst and Robert W. Heath Jr. are with North Carolina State University, Raleigh, NC 27695 (rmdreifu@ncsu.edu, rwheathjr@ncsu.edu).
			This material is based upon work supported in part by the National Science Foundation under Grant No. NSF-ECCS-2153698, NSF-CCF-2225555, and NSF-CNS-2147955}
	}
	
	\maketitle
	\bstctlcite{IEEEexample:BSTcontrol}
	
	\section*{Abstract}
	Massive multiple-input multiple-output (MIMO) is an important technology in fifth generation (5G) cellular networks and beyond. To help design the beamforming at the base station, 5G has introduced new support in the form of flexible feedback and configurable antenna array geometries. 
	In this article, we present an overview of MIMO throughout the mobile standards, highlight the new beam-based feedback system in 5G NR, and describe how this feedback system enables massive MIMO through beam management. Finally, we conclude with challenges related to massive MIMO in 5G.

	\section*{Introduction}
	Multiple-input-multiple-output (MIMO) is a general class of technologies that incorporates a number of transmission and reception techniques using multiple antennas. Spatial multiplexing (SM), among the most well known MIMO methods, involves the transmission of multiple data streams, called layers in 3GPP specifications. In a single user (SU-)MIMO setting, spatial multiplexing involves directing multiple streams to one user, resulting in an increase in the spectral efficiency proportional to the number of streams. Similarly, the concept can be applied in a multiple user (MU-)MIMO setting, where the layers may be split among users.

\begin{figure*}[!t]
		\centering
		\includegraphics[width=6.in]{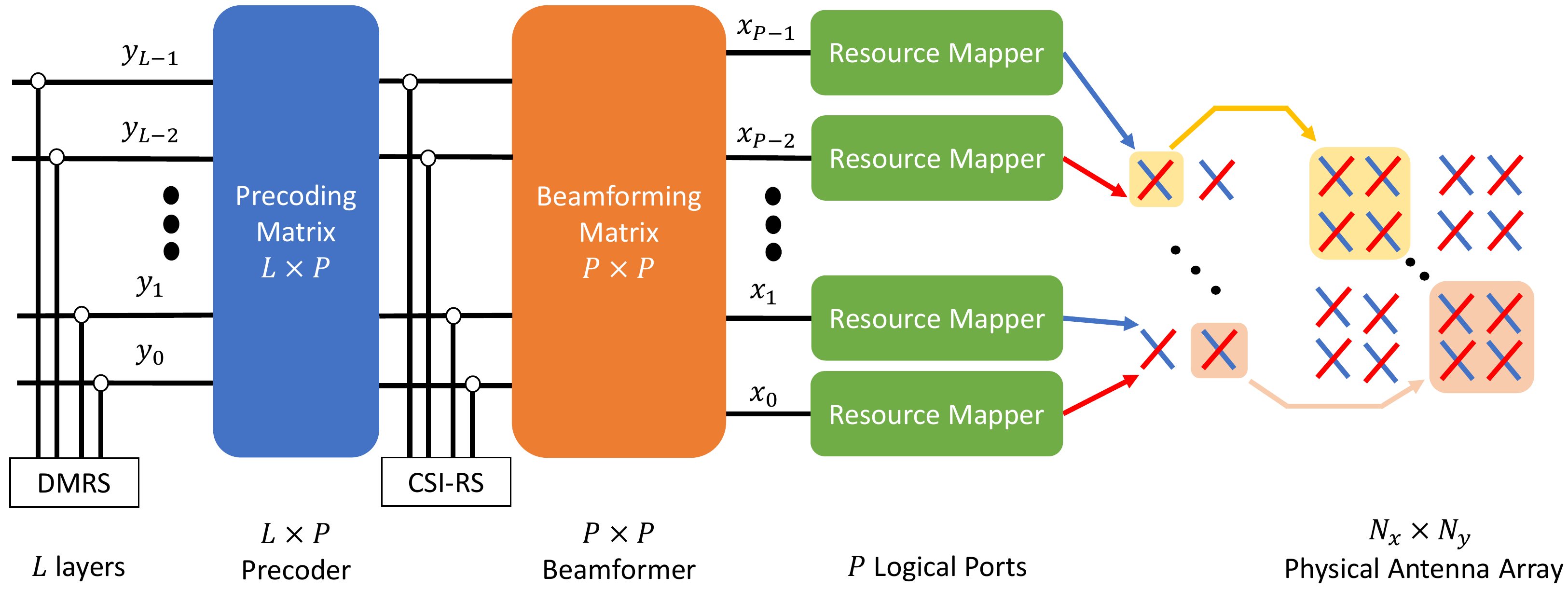}
		\caption{The data processing flow from data layers to antenna outputs in a 5G base station. First the streams (and DMRS) are multiplexed according to the precoding matrix, which assigns the layers to the ports. Then, the result, along with CSI-RS, is beamformed to the logical ports. The logical processes are then mapped onto OFDM resource grids and potentially beamformed again onto the corresponding physical antennas. 5G uses dual polarized arrays to transmit multiple layers on orthogonal electromagnetic wave directions.}
		\label{fig: ports}
	\end{figure*}

	
	MIMO usage in mobile standards has played a growing role since the inception of SM in 3GPP Release 7. MIMO was a backbone of the evolved high speed packet access (HSPA+) that would enable doubling the achievable data rates with two-layer MIMO. The subsequent Release $8$ would increase the downlink MIMO capabilities to $2\times2$ with two layers each going to two users, although uplink capabilities were still limited to a single user. Release $8$ would also introduce other forms of multi-antenna techniques in the form of \textit{transmission modes}. These formats include single-antenna, transmit diversity, open-loop SU-MIMO, closed-loop SU-MIMO, closed-loop rank-$1$ precoding (beamforming), and MU-MIMO. Releases $9$ and $10$ would introduce larger array sizes--up to $8$ downlink antennas--and transmission modes $8$ and $9$ which extend the closed-loop SM to $8$ antennas or transmit diversity if feedback is unavailable. The introduction of larger arrays also enabled multi-layer beamforming (BF), which is different from previous SM by using multiple antennas for each layer. Academically, this is usually referred to as \textit{precoding}, but precoding is used in 3GPP notation to describe the multiplexing of the data streams onto the ports as shown in Figure \ref{fig: ports}. 
	
	Throughout Releases $8-12$, antenna arrays have been assumed to have the same structure where antennas are built in columns and each column (as well as each polarization in dual-polarized arrays) is separately controlled for SM in the azimuth direction. Some basic degree of elevation control was further enabled with mechanical and electrical tilting of the arrays, though it was not until later that tilting was dynamically controlled and became standardized. Release $13$ introduced a study item on full-dimension MIMO (FD-MIMO) that included controllable elements in both the azimuth and elevation directions for 3D beamforming. Additionally, the mapping of ports and physical antennas was separated so that multiple antennas could be controlled from a single channel state information (CSI) port (e.g. there does not need to be a one-to-one mapping). This paradigm shift in antenna arrays allowed networks to use larger, more directive arrays without needing to update the standards or feedback by ``grouping'' antennas into subarrays that are transparent to the UE. With the interest in larger arrays and separation of physical antennas and logical elements, Release $13$ can be seen as a pivotal point in the evolution towards massive MIMO (M-MIMO).

	{\renewcommand{\arraystretch}{1.5}
		\begin{table*}[!t]
			\caption{3GPP jargon summary} \label{tb: acronyms}
			\centering
			\begin{tabular}{|c|c|}
				\hline
				\textbf{Term} & \textbf{Description} \\ \hline
				MIMO             &               Multiple-input-multiple-output technologies \\ \hline
				Layers           &               Data streams using the same time-frequency resources \\ \hline 
				TM               &               Transmission mode; defines a set of supported MIMO techniques like SM, beamforming, etc. \\ \hline
				SM               &               Spatial multiplexing; MIMO method of transmitting multiple data layers \\ \hline
				FD-MIMO          &               A MIMO layout with beamforming in azimuth and elevation direction \\ \hline
				Port             &               A non-unique subset of antenna elements controlled by an RF chain \\ \hline
				RSRP             &               Reference signal received power \\ \hline
				Feedback         &               A general term encompassing the information from the RX based on reference signals \\ \hline
				CSI              &               Channel state information; partial or complete knowledge of the wireless channel \\ \hline
				RI               &               Rank indicator; provides maximum supportable layers as feedback \\ \hline
				CQI              &               Channel quality indicator; encodes the reference signal metric (i.e. RSRP, RSSI) \\ \hline
				PMI              &               Precoder matrix indicator; provides feedback for closed-loop MIMO \\ \hline
				SSB              & Synchronization Signal Block; for coarse beam training, synchronization, and initial access \\ \hline
				CSI-RS           &               CSI reference signal; a known pilot sequence used for beam training and channel estimation \\ \hline
				DMRS             &               Demodulation reference signal; a known pilot sequence used to aid demodulation \\ \hline
				
			\end{tabular}
		\end{table*}
	}

	{\renewcommand{\arraystretch}{1.5}
		\begin{table*}[!t]
			\caption{Comparison of flexible configurations in LTE and 5G NR} \label{tb: LTE5G}
			\centering
			\begin{tabular}{|c|c|c|c|c|}
				\hline
				& \textbf{Subcarrier Spacing} & \textbf{MU-MIMO Support} & \textbf{CSI-RS Ports} & \textbf{MIMO Feature} \\ \hline
				\rowcolor[HTML]{ECF4FF} 
				\textbf{LTE} (Rel $8$-$9$)                 & $15$ kHz                      & $4\times2$                  & $\{1,...16\}$                        & TM $1$-$8$    \\ \hline
				\textbf{LTE-A} (Rel $10$-$12$)                 & $15$ kHz                      & $4\times2$                  & $\{1,...16\}$                        & Codebook BF \\ \hline
				\rowcolor[HTML]{ECF4FF} 
				\textbf{LTE-A Pro} (Rel $13$-$14$)                 & $15$ kHz                      & $8\times2 \text{ or } 4\times4$                & $\{1,...32\}$                        & FD-MIMO \\ \hline
				
				\textbf{5G FR1 {[}0-3GHz{]}} (Rel 15-17)& $\{15, 30\}$ kHz              & $8\times2 \text{ or } 4\times4$                  & $\{1,...32\}$                        & $4$ beam SSB \\ \hline
				\rowcolor[HTML]{ECF4FF} 
				\textbf{5G FR1 {[}3-6GHz{]}} (Rel 15-17) & $\{15, 30, 60\}$ kHz          & $8\times2 \text{ or } 4\times4$                  & $\{1,...32\}$                        & $8$ beam SSB \\ \hline    
				\textbf{5G FR2} (Rel 15-17)             & $\{60, 120, 240\}$ kHz        & $8\times2 \text{ or } 4\times4$                 & $\{1,...32\}$                        & $64$ beam SSB  \\ \hline
			\end{tabular}
		\end{table*}
	}

\begin{figure*}[!t]
		\centering
		\includegraphics[width=6.in]{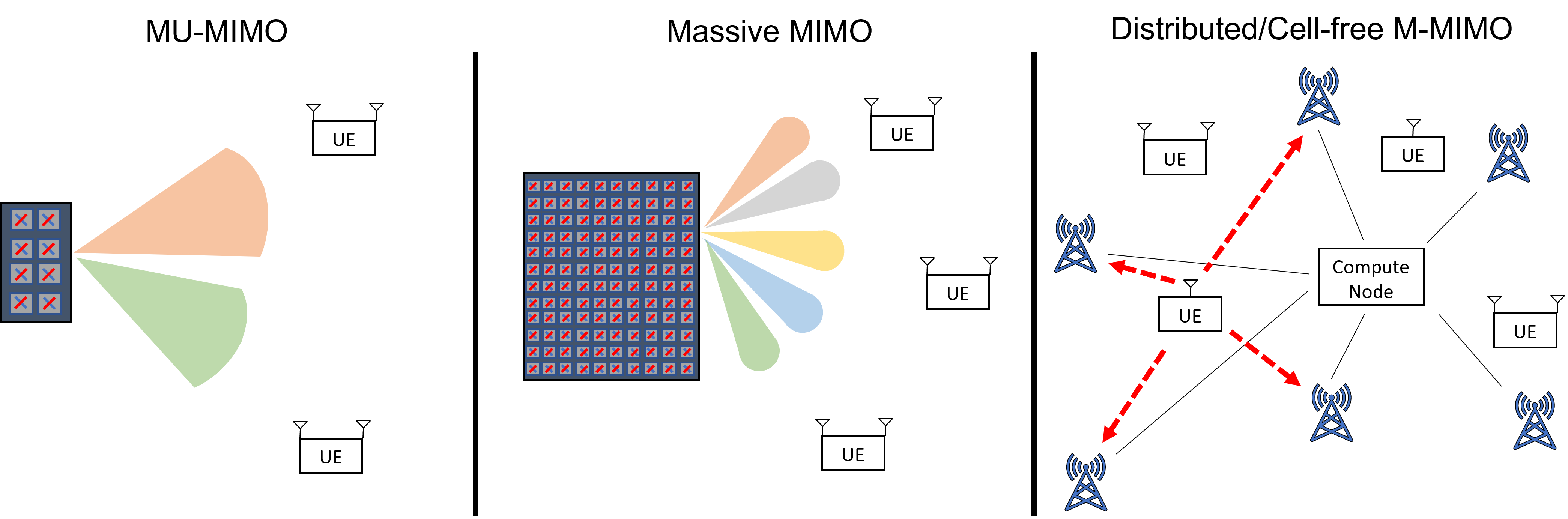}
		\caption{Visualization of MU-MIMO, massive MIMO, and coordinated multi-point or cell-free M-MIMO. In cell-free massive MIMO, there is a centralized compute node, multiple access points, and user equipment (UE). The compute node uses the distributed access points like logical elements and processes the data across the network.}
		\label{fig: cellfreemimo}
	\end{figure*}

	Terminology in mobile broadband includes many acronyms and differences from academic wireless research. To assist readers, Table \ref{tb: acronyms} outlines important terms necessary for understanding MIMO and feedback in 3GPP standards. An important distinction in mobile networks is an additional set of terms to differentiate the physical antennas from the logical processing chains and further-up data layers, as shown in Figure \ref{fig: ports}. The data layers are each associated with demodulation reference signals (DMRS) for estimating the effective channel seen by the user equipment (UE). The layers are then precoded over logical antenna ports, which are each mapped to individual resource grids and then sent to the physical antennas. Furthermore, polarization naturally applies to the system by mapping two logical ports onto orthogonal polarized antennas. The logical ports can apply to a non-exclusive subset of the physical antennas, so long as the same subset is used consistently. Each logical port is equipped with its own resource grid and mapper to take advantage of the flexible delineation between ports and physical antennas. Further evolution of the flexibility occurs when the physical antennas are not co-located, which enables multi-panel arrays and cell-free massive MIMO. The coordination of disaggregated antenna panels was introduced in LTE-A release $11$ as a centralized radio access network (C-RAN) although the idea has been brought back up in recent years as distributed MIMO or cell-free massive MIMO. We provide a simple visualization of these situations in Figure \ref{fig: cellfreemimo}.
	
	One of the cornerstones of 5G is a focus on flexibility. To that end, 5G NR incorporates a flexible numerology, configurable bandwidth parts, and numerous feedback formats. This flexibility is especially important for the two frequency ranges (FR1: $0$-$6$GHz, FR2: $\ge 6$GHz) which correspond to different subcarrier spacings, bandwidths, and feedback regularity. We summarize some of the possible configurations in Table \ref{tb: LTE5G}. Of the changes in 5G, the new format beam management based on synchronization signal blocks (SSB) and channel state information reference signals (CSI-RS) is distinctly different from the CSI feedback process in LTE. 
	The primary reasons for the new processes are: 1) Beamforming is enabled at every step including initial access to improve the SNR during synchronization and channel estimation. 2) Beam management integrates analog beam training with CSI feedback for new antenna array configurations like hybrid architectures. 3) The new feedback format (type-II) provides high resolution CSI feedback to improve MU-MIMO performance.
	
	In the next section, we will describe the reference signals, which include known pilot sequences and configuration information, used in the downlink for 5G networks. There are additional reference signals for uplink-based beam management, but most of the focus will be on the downlink herein. In the following sections, we will describe the feedback associated with beam management and address how the integration of beam training improves the initial access performance and enables new, massive architectures. 
	
	\section*{Reference signals} \label{sec: RSRP}
	Reference signals (both SSB and CSI-RS) serve important tasks like synchronization, channel estimation, and handover in 5G networks. Within the reference signals, there are known pilot sequences with mathematical properties that aid tasks like synchronization, as well as configuration information like the base station capabilities and logical geometry. 5G NR uses synchronization signal (SS) blocks, as a reference signal for synchronizing user equipment (UE) and obtaining basic feedback during initial access. SSBs are transmitted periodically by the base station once every $\{5, 10, 20, 40, 80, 160\}$ ms. The base station transmits a specific primary and secondary synchronization sequence as well as embedded DMRS. In total, the SSB waveform spans $240$ subcarriers and $4$ OFDM symbols for each block. The entire process includes at least one but as many as $\{4, 8, 64\}$ SS Blocks in an SS-Burst, depending on the frequency range, each of which is transmitted time-sequentially according to \cite[Section 4.1]{3gppTS38213}. Note that both frequency and time offsets can be set to prevent SSB collisions with nearby stations. By using multiple SSBs, the base station can beamform each one differently so that the UE can listen to the entire burst and provide the BS with the best beamformer index within the measurement report, thereby selecting a potential beam for downlink transmission or angular direction for requesting additional feedback.
	
	In contrast to SSB reference signals, CSI-RS can be configured with reference symbols transmitted across a wider bandwidth and can occur periodically or aperiodically. CSI-RS also enable feedback for massive hybrid arrays by training the analog beamforming with precise CSI-RS beamformers without the UE needing knowledge of the physical antenna geometry. The CSI-RS index defines the UE-recommended analog beamformer and is always fed back in every feedback packet. If PMI is configured, the UE must also provide precoding information using knowledge of the logical port geometry provided by the BS. The drawback to CSI-RS, especially for large arrays, is an excess of overhead and potentially out-of-date information. In particular, UEs are not required to update the CSI-RS index for a CSI-RS resource if the process has been updated within the last $5$ subframes, or if the number of CSI processes exceeds the limitations defined in Table 7.2.1 of \cite{3gppTS38213}. This means that channel information may become stale for large arrays which leads to misaligned beamforming and suboptimal performance. Furthermore, CSI-RS are expensive to allocate because only one CSI-RS is utilized over a set of resources to prevent interference.
	
	\section*{Feedback} \label{sec: feedback}
	The base station can configure CSI reports, which are packets containing feedback, either periodically or aperiodically, with reference signals multiplexed between the ports. A report generally includes the channel quality indicator (CQI) and a reference signal indicator (SSBRI/CRI). Additionally, a rank indicator for multi-stream communication and a precoding matrix indicator (PMI) can be included in the CSI report. Furthermore, some quantities such as CQI and PMI can include both wideband (average across the bandwidth) and subband components to increase the feedback and precoder responsiveness in frequency selective channels. The CQI provides a measure of the strength of the channel and is used to determine the modulation order and code rate for the downlink transmission. The reference signal indicator is used to report the strongest received reference signal index of the beamformed reference signals for beam training. The PMI field has different characteristics depending on the feedback format but it carries the primary CSI information for the base station. 
	
\begin{figure}[!t]
		\centering
		\includegraphics[width=3.45in]{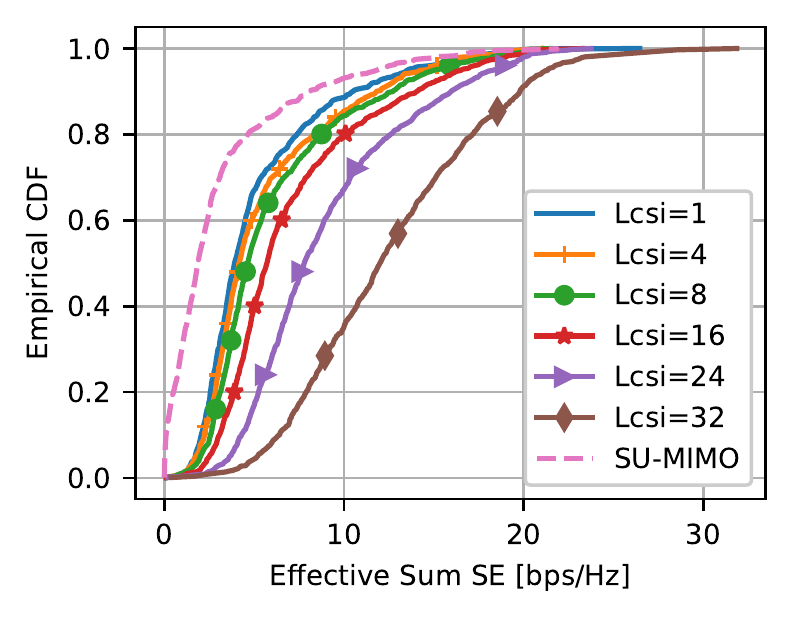}
		\caption{A comparison from our work \cite{Dreifuerst2023CodebookFeedback} of the effective sum spectral efficiency (SE) achieved in a simulated FR1 deployment using CSI type-II feedback with $L_{\text{CSI}}$ multi-path feedback quantization. It can be seen that sub-6GHz environments have very rich scattering that cannot be captured effectively with the current 5G feedback limitations of $L_{\text{CSI}} \le 6$ for large MIMO arrays.}
		\label{fig: feedback_limit}
	\end{figure}

	There are two formats for PMI: predefined (type-I) or constructed (type-II). In LTE, the PMI format is always predefined, meaning the UE would select the PMI from an established table of precoder combinations according to a metric, e.g. maximizing signal-to-interference-noise ratio (SINR). The benefit of type-I PMI is that it requires low overhead and is computationally simpler for the UE to calculate. In contrast, type-II PMI is more flexible and precise but it comes at the cost of higher computational complexity and larger overhead. Constructed PMI is built up as a sum of $\Lcsi$ multipath components, which are represented by oversampled 2D DFT beamforming vectors, with quantized amplitude and phase components. An iterative process is used in combination with a metric like SINR to determine the beamforming vectors and complex weights, although, the oversample DFT is reduced to an orthogonal basis after the first multipath component is selected so that the $\Lcsi-1$ beams are only chosen from the remaining orthogonal DFT vectors. While type-II feedback is able to more accurately quantize the CSI, it is still limited by the channel estimation accuracy and the number of multipath components supported in the specification. Release $16$ is currently restricted to at most $\Lcsi=4$ in type-II feedback and $\Lcsi=6$ in type-II enhanced feedback. This is a strict limitation in rich scattering environments such as macro-cell FR1 deployments, as seen in Figure \ref{fig: feedback_limit}. Channel estimation, however, can be improved through better estimators or by increasing the RSRP through beamformed pilot signals.

	\section*{Beam Management and Massive MIMO} \label{sec: beam_management}
	Beam management is designed to unify the reference signals, channel state information, and feedback into one process \cite{Giordani2019}. At the heart of the process are three steps: initial access, beam reporting, and beam refinement. In some cases, the process is expanded to include beam tracking for mobile UEs or with separate refinement stages for the base station and UE. Ultimately, beam management is built around the flexibility of the feedback system in 5G NR. 
	
	The initial access period includes the transmission of beamformed SSBs that provide the UEs with a basic synchronization signal and demodulation reference signals. This allows for UEs to save power by going inactive and rejoining the network at a later initial access period. At the physical layer, the UE will receive the SSBs from a single antenna or using one or more spatial filters, such as a multi-panel handset used to overcome hand blockage. The UE will use the received SSB for synchronization and determining the control information.
	The beam reporting stage includes one or more possible SSB CSI reports which are transmitted in the random access channel. The report includes information for the strongest serving cell and may include a set of the next strongest cells within the same band to assist with load balancing. The number of reported additional cells depends on the carrier frequency, the previous state of the UE in the network, and the bands being monitored. In a newly-active state, the UE reports the top $6$-$16$ additional cells across each active frequency range \cite{3gppTS38213}. This reporting helps to manage handover and mitigate cell-edge interference. 
	In the final steps, the UE has connected to a serving cell and is ready to start receiving data. Further beam refinement and channel estimation can occur by transmitting reference signals with more precise beams. Although not specified in the standard, a typical CSI-RS would cover smaller portions of the reported SSBs' directions or combine coherently across a multipath channel. Using more directional or precise beams can increase the SNR--thereby improving the channel estimates and beam alignment. Beam refinement can also be used to adjust the beamforming slightly to track highly mobile UEs.
	
	One of the key limitations of the beam management framework is the finite resources available for sweeping, refinement, and tracking. For example, the SSB sweeping process is limited to $\{4, 8, 64\}$ SSB beams, which are broadcast to all users, and each beam is restricted to just $240$ subcarriers and $4$ symbols. This corresponds to $5\%$ or less of the resources available for downlink data transmission. In contrast, CSI-RS can cover an entire bandwidth part, typically assigned as user-specific, and up to $32$ ports can be configured for CSI-RS. The port limitation is an important one because it defines the maximum dimension that the UE will support for channel estimation and PMI selection. In order to support larger arrays, Base stations are deployed with a combination of directly-connected antennas and hybrid arrays with phase shifters connecting the $32$ CSI-RS ports to a set of antennas. The analog portion of the hybrid array can be trained with the SSB and CSI-RS beams, while the PMI can be used to determine the digital precoder. The integrated beam training and CSI acquisition in 5G enables a new level of support for arbitrary, massive arrays.

	\section*{FR1 and FR2}
	The key differentiator between FR1 and FR2 i.e. sub-6GHz compared to mmWave--for a set of antenna arrays of equal aperture--is the propagation environment, bandwidth available, and reliance on beam-based architectures. At sub-6 GHz, the propagation environment tends to be more reflective. This results in multipath propagation that is beneficial for traditional spatial multiplexing. In FR1, users are not expected to beamform, so beam training is only necessary on the BS side. Furthermore, because BS arrays have a limited physical size and therefore a relatively small number of FR1 antennas, the beam-based system is used in a limited format \cite{SamsungWhitePaper2020}. For example, the UE may only employ a single antenna during SSB reception and uplink transmission and the BS may reduce the number of active SSB beams because precise beamforming is unnecessary in low-band environments. 
	
	In FR2, the full beam-based system is employed with BS beam training, repetition, and UE beam training all occurring. During the refinement period, the BS can sweep over a subset of the CSI-RS codebook, and feedback is gathered via one or more CSI-RS measurement reports. When increasing the carrier frequency to millimeter-wave bands, the paths tend to have a strong line of sight component, but very weak reflected components that restrict multi-layer communication.
	In fact, MU-MIMO does not appear to be active in any FR2 developments, and even $4\times2$ MU-MIMO in FR1 has only recently been recorded in commercial networks \cite{SignalsFlash2022MIMO0}. While FR1 deployments are often able to support more spatial multiplexing layers, the bandwidth of a mmWave channel in FR2 is much larger, allowing for as much as four times larger data rates.

	\section*{Challenges} \label{sec: Challenges}
	There are a series of challenges associated with massive MIMO compared to traditional architectures. Questions of power consumption, feedback, hardware cost, computational/algorithmic complexity, and robustness are all exacerbated with massive arrays. Furthermore, these challenges are also impacted by the specifications and beam management framework in particular. Here we outline $3$ challenges specific to M-MIMO arrays in the 5G beam management framework.
	
	First, prior to any deployment, a group of codebooks must be designed. The design of codebooks is critical due to varying RF environments and user mobility patterns. In particular, a BS must have codebooks for: 1) initial access (SSB) coverage, 2) refinement (CSI-RS), and 3) feedback and mobility. For case 1, the codebook is generally small due to the limited number of SS blocks defined by $N_\text{SSB}$ in Table \ref{tb: LTE5G}, which is never more than $64$. Furthermore, the SSB process has a short repetition period with a default of $20$ms. In contrast, the CSI-RS codebook used for refinement is often much larger to maximize the signal quality and alignment of the beam with a user. For example, the CSI-RS codebook might contain as many as $4-16$x more beams than the number of BS RF chains. The BS would make use of these beams at a slower timescale than the SSB process, typically $80$ms with additional aperiodic usage as needed. While both of these codebooks can be specific for a given BS, the third codebook that is used for feedback and mobility must be known by the entire network. The feedback codebook enables quantizing the multipath information at the UE side and providing the quantized representation in the PMI for the BS. In the case of reciprocity (e.g. in time-division duplexing), the PMI may be reduced or neglected, although the feedback is often still helpful due to the wireless link being asymmetric \cite{Bjornsson2022CombiningCSIFeedback, SamsungWhitePaper2020}.
	Therefore, the UE and the BS must both use the same codebook for feedback to correctly share the PMI information. The design of codebooks for SSB and CSI-RS is an active area of research due to the significant gains that can be achieved over generic strategies like DFT codebooks \cite{R1.166222}. Additionally, although enhanced type-II feedback codebooks have already been implemented in 3GPP release $16$, MU-MIMO performance is still severely limited and new codebooks that can efficiently support massive MIMO arrays are needed to reduce the performance loss.

	Once the codebooks are determined, an intelligent process for beam selection and sweeping is necessary. The process of sweeping and beam selection has generally received the most focus in beam management tasks \cite{Sim2018MABBMV2X, Wangetal2009CodebookTraining, HeathetAlVehicularSensing2016}, although it is still largely unclear how system performance is impacted by new sweeping algorithms. In particular, characterizing when the improved alignment outweighs the cost of additional overhead and interference in realistic, large-scale settings is still an active area of research. The greatest reduction of overhead could be seen by minimizing the CSI-RS usage because a typical CSI-RS codebook can include hundreds or thousands of beamforming vectors in modern 5G codebooks. At the same time, CSI-RS beam selection is heavily dependent on the SSB beam selection. Some of the simplest sweeping algorithms are based on hierarchical or tiered search \cite{Wangetal2009CodebookTraining, Morozov2016}, where all of the CSI-RS beams within an SSB beam are used for each user. In an M-MIMO array the angular resolution is very fine, though, so there are many possible CSI-RS beamformers within an SSB beam and the difference in SNR between the beams could be significant depending on the environment. Other algorithms have been proposed based on machine learning, compressive sensing, and channel statistics \cite{Sim2018MABBMV2X} and references therein. 3GPP has also introduced a study item on machine learning with an explicit focus on improving beam management and feedback through artificial intelligence. These works present new potential directions, but significant research still remains in evaluating such algorithms in realistic scenarios.
	
	Finally, a critical issue in beam management is the challenge of mobility robustness. Even with sufficient feedback at one time instance, the BS needs to update the information at least as fast as the channel coherence time to accurately direct the beams. The coherence time, though, depends on the carrier frequency and mobility \cite[section 3.4.3]{HeathLozano2018}. Furthermore, with highly directive beamforming, mobility can result in misaligned beams that reduce performance or even cause radio link failure \cite{HeathetAlVehicularSensing2016}. Mobility is especially challenging for vehicular UEs in FR2 bands, which have mobility patterns and extremely directive beams that must adjust frequently. While mobility is challenging for the base station, it is often more difficult for a UE. The UE needs to update the combining weights to match the beamformed channel coherence time, which is on the order of milliseconds. To assist with this, multi-panel arrays \cite{rebato2018multiTRP} have been standardized in Release 16 to reduce link failure and improve robustness. Multi-panel and geometry-aware research is expected to become an active area of interest as a result of the recent standardization. Research efforts have also attempted to improve mobility management with methods such as multi-modal data \cite{HeathetAlVehicularSensing2016} and machine learning \cite{Hussain2022DualtimeBeamTracking}. These methods tend to consider single-beam or single-layer data in mmWave settings, but spatial multiplexing will further challenge mobile communications due to the precise precoding and combining required to achieve coherent processing.

	\section*{Conclusion}
	In this article, we have brought together the ideas of feedback and massive MIMO in light of the recent releases of 5G NR. 
	The first focus has been on the integration of beam management, reference signals, and feedback that are enabling MU-MIMO in network deployments. Still, both FR1 and FR2 deployments have yet to reach the potential for massive MIMO due to challenges like codebook design and mobility robustness. We expect massive MIMO will be an active area of research and industrial growth throughout the continued development of 5G and future wireless generations.
	
	\bibliographystyle{IEEEtran}
	{\footnotesize
		\bibliography{IEEEabrv, references, heath_refs_all}
		\begin{IEEEbiography}{Ryan Dreifuerst}
			is a PhD student at North Carolina State University. He obtained his B.S. in Electrical Engineering from Milwaukee School of Engineering with minors in Mathematics and Physics. He also received his B.S. in Electrical and Communications Engineering from Technische Hochschule L{\"u}beck. He received his M.S. in Electrical Engineering at The University of Texas at Austin (UT Austin).
			Ryan has spent the previous summers working with Meta, Qualcomm, and Samsung Research America on 5G and 6G wireless systems.
			His research lies at the intersection of machine learning and signal processing. Specifically, he has focused on augmenting machine learning with domain knowledge for physical layer processing and beam-based MIMO algorithms.
		\end{IEEEbiography}
		\begin{IEEEbiography}{Robert W. Heath Jr.}
			is the Lampe Distinguished Professor in the Department of ECE at North Carolina State University. He is the recipient or co-recipient of several awards including  the 2019 IEEE Kiyo Tomiyasu Award, the 2020 IEEE Signal Processing Society Donald G. Fink Overview Paper Award, the 2020 North Carolina State University Innovator of the Year Award, and the 2021 IEEE Vehicular Technology Society James Evans Avant Garde Award. He authored "Introduction to Wireless Digital Communication” (Prentice Hall in 2017) and "Digital Wireless Communication: Physical Layer Exploration Lab Using the NI USRP” (National Technology and Science Press in 2012). He co-authored “Millimeter Wave Wireless Communications” (Prentice Hall in 2014) and "Foundations of MIMO Communications" (Cambridge 2019). He is a licensed Amateur Radio Operator, a registered Professional Engineer in Texas, a Private Pilot, a Fellow of the National Academy of Inventors, and a Fellow of the IEEE.
		\end{IEEEbiography}
	}
\end{document}